\documentclass[shortAfour,sageh,times]{sagej}
\usepackage{mathtools, cuted}
\usepackage{moreverb,url}
\usepackage{graphicx}
\usepackage{dcolumn}
\usepackage{amsfonts}
\usepackage{amssymb}
\usepackage{amsmath}
\usepackage{slashbox}
\usepackage[colorlinks,bookmarksopen,bookmarksnumbered,citecolor=red,urlcolor=red]{hyperref}

\newcommand\BibTeX{{\rmfamily B\kern-.05em \textsc{i\kern-.025em b}\kern-.08em
T\kern-.1667em\lower.7ex\hbox{E}\kern-.125emX}}

\def\volumeyear{2016}

\begin{document}

\runninghead{Nana et \textit{al.}}

\title{Effect of tendon structural control on the appearance of  horseshoes chaos on a cantilever beam due to seismic action}

\author{B.R. Nana Nbendjo$^{1,\ast}$ and U. Dorka$^{2}$}

\affiliation{$^{1}$Laboratory of Modelling and Simulation in Engineering and Biomimetics and  Prototypes, University of Yaound\'{e} I, P. O. Box 812, Yaound\'{e}, Cameroon.\\ $^{2}$Steel and Composite structures, Kassel University, Kurt-Wolters-strasse 3,Kassel 34125, Germany.\\$^{\ast}$Corresponding Author: nananbendjo@yahoo.com (B.R. Nana Nbendjo)}



\begin{abstract}
 This paper deals with the problem of improving the seismic
strength of mechanical structures by using  a tendon system as
actuation device. It consists of a pair of tension cables
transmitting a control torque to the structure at the moment arm
position. The purpose of the paper is to, establish the analytical
framework consisting of mathematical modelling of cantilever beam
under active tendon structural control, analyze the stability and,
determine the physical characteristics of tendon system leading to
the suppression of horseshoes chaos. The control efficiency is
found by analyzing the behavior of the controlled system through
the Melnikov methods. We also provide the critical control gain
parameters leading to the efficiency of the control process.
\end{abstract}


\maketitle

\section{Introduction}
For years, the problem of seismic induced vibrations in mechanical
structures has received broad attentions. It has now reached the
stage where active systems have been installed in full-scale
structures \cite{R1,R2}. This paper demonstrates theoretically a
possibility of implementation of active tendon control to a
full-scale structure under actual ground motions. One method used
to handle this problem is the installation of automatic active
control forces in mechanical structures as mentioned by \cite{R3}.
Numbers of structural concepts have been identified \cite{R4}
which allow rigid body control and four concepts (Base Isolation,
Hysteretic Device System, Tendon system and Pagoda system) have
been suggested for seismic control \cite{R5}. Among those, active
control using tendons has been one of the most promising technics
to implement control mechanisms \cite{R6,R7,R8,R9}. These systems
consist of a set of prestressed tendons connected to a structure
where a servomechanism controls their tensions. The tendon is
generally modeled as a simple spring or as a spring and dashpot in
parallel.  This simplicity makes it attractive and also helpful
for
retrofitting or strengthening an existing structure.\\
${\quad}$${\quad}$ Figure 1 illustrates the idea of a tendon
control system for suppressing cantilever beam vibrations. A pair
of actuators at the beam root activates the tendons (\emph{i.e.},
tension cables) to rotate a pair of moment arms attached at a
proper position of the structure. Thereby, the beam motion is
actively controlled by using the feedback signals from  sensing
devices located on top  of the beam. The tendon control is
suitable to this task not only because the hardware is simple to
implement, but also because a robust colocated control is realized
by using "non-
colocated" sensors and actuators \cite{R8,R9}.\\
${\quad}$${\quad}$ In this paper, we are interested by the
condition for which tendon system could avoid the global
bifurcation before and after loss of stability \cite{R10,R11}.
Since these conditions can be detected by means of basin of
attraction, it is important to obtain the criteria for
theoretically quenching chaotic dynamics. This may imply the
existence of fractal basin boundaries and the so-called horseshoes
structure of chaos. It will be shown that, under suitable
hypotheses, chaos arises quite naturally in the dynamics of the
cantilever beam subjected to seismic action. Chaotic motions of a
beam were earlier studied by  \cite{R12}, and a general review of
the application of the theory of chaos to elastic structures can
be found in some previous works \cite{R13,R14}. Melnikov method
developed by Guckenheimer and Holmes \cite{R15}, is employed for
chaos detection in the system. This corresponds to
the occurrence of transverse heteroclinic intersections.\\
${\quad}$${\quad}$ The cantilever beam model with tendon system
control is illustrated in Section 2, where the exact equation of
planar motion is also obtained. In Section 3, we analyze static
bifurcation behavior, which involves transverse intersection
between perturbed and unperturbed separatrix. Therefore the
critical parameters  of the tendon system under which bifurcation
is suppressed are pointed out. Section 4 is devoted to conclusion.

 \begin{figure}[!h]
   \centerline{\fbox{\vbox {\hsize0.1cm\hrule width0.1cm height0pt}\includegraphics[width=3.0in]{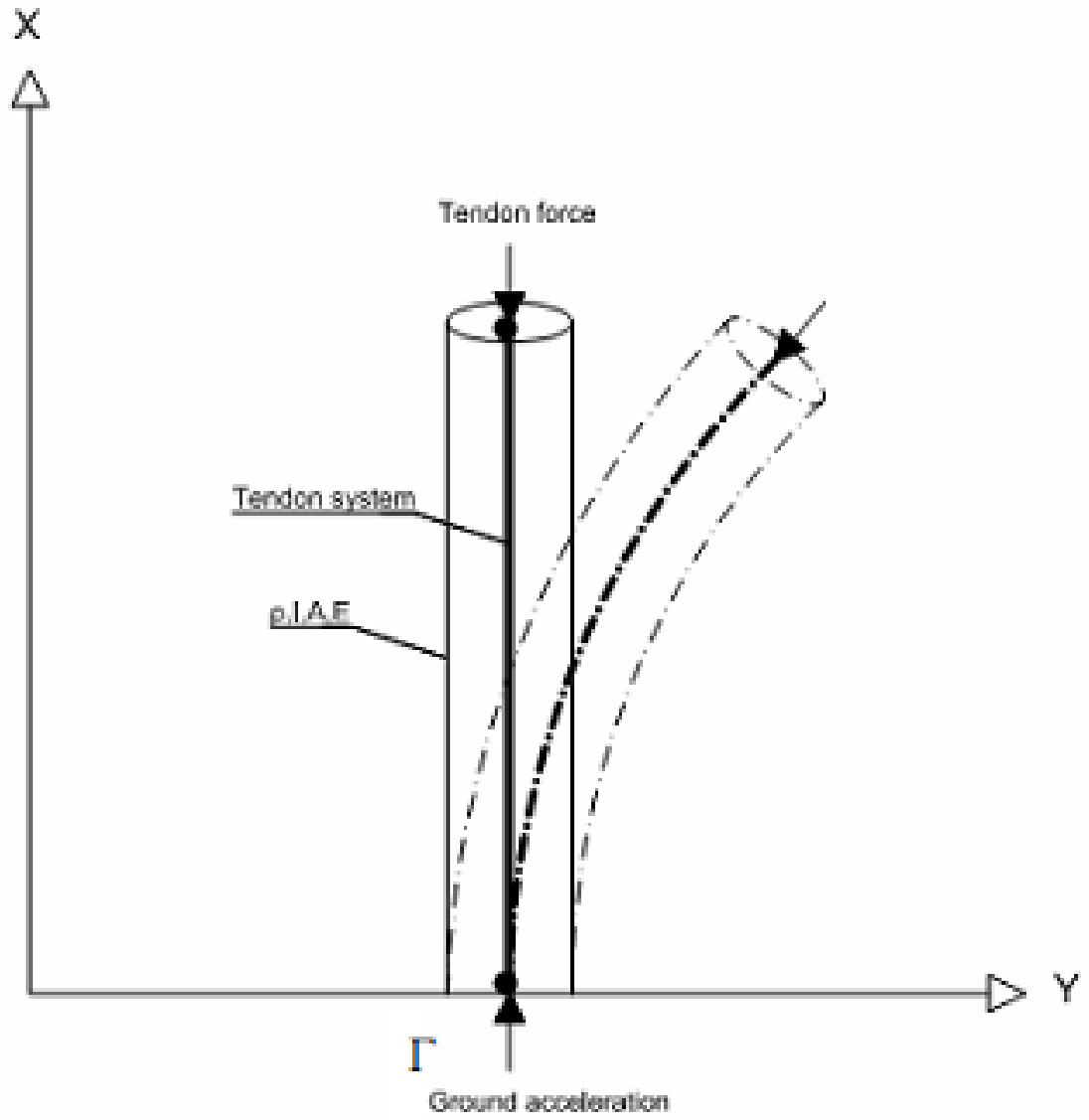}}}
  \caption{Tendon system concept on a cantilever beam}\label{fig.1}
 \end{figure}
\section{General mathematical formalism of a cantilever beam under tendon system control subjected to seismic action}
 ${\quad}$${\quad}$Consider a cantilever beam of length $l$, with density
$\rho$, Young modulus $E$, cross sectional area $S$ and  moment of
inertia $I$. The governing equation for its vibration under active
tendon structural control (see figure 1) subjected to seismic
action obtained by standard methods is given by
\begin{eqnarray}
\label{eq1} &&\rho S\frac{\partial^{2}y}{\partial
t^{2}}+EI\frac{\partial^{4}y}{\partial x^{4}}+\lambda
\frac{\partial y}{\partial
t}+\Gamma(t)\frac{\partial^{2}y}{\partial
x^{2}}\nonumber\\&&-k[\int_0^{l}(\frac{\partial y}{\partial
x})^{2}dx]\frac{\partial^{2}y}{\partial x^{2}}=z_c\delta(x-l),
\end{eqnarray}
where $y=y(x,t)$ is the lateral deflexion. $\lambda$ represents
the viscous damping, $k$  the nonlinear stiffness, and $\Gamma(t)$
is the ground acceleration applied to the beam. The nonlinear term
expresses the fact that the axial force in the beam increases with
lateral deflexion, leading to an increase of restoring forces.
$z_c$ is the control force and $\delta(x-l)$ materializes the fact
that this force act on top of the beam. The control force is given
by \cite{R7}
\begin{eqnarray}
\label{eq2}
z_c(t)=4k_{c}cos{\alpha_{c}}[s_1y(t-t_x)+s_2\frac{\partial
y}{\partial t}(t-t_{\dot{x}})],
\end{eqnarray}
where $k_c$ is the tendon stiffness, $\alpha_c$ the tendon
inclination, $s_1$ and $s_2$ the control gain parameters. $t_x$
and $t_{\dot{x}}$ are time delays for displacement and velocity
feedback force in the system respectively. We notice that $x$ and
$\dot{x}$ are written as a function of $t-t_{x}$ and
$t-t_{\dot{x}}$ respectively.  In fact tendons are viscoelastic
structures which means they exhibit both elastic and viscous
behaviors. When stretched,  the stress-strain curve
 starts with a very low stiffness region, and then  the structure becomes significantly
 stiffer, and behaves reasonably  until it begins to fail.\textbf{ The tendon can be  usually installed by drilling holes
 into existing walls or columns and is anchored in both ends.
 One of these ends may present advanced control mechanisms, like (non-linear)
  springs or shape-memory-alloy devices}. \\ ${\quad}$${\quad}$
The cantilever beam is fixed at its base $(x=0)$ and free at the
top ($x=l$), which implies the following boundaries conditions:
\begin{eqnarray}
\label{eq3} y(0,t)=\frac{\partial y}{\partial
x}(0,t)=\frac{\partial^{2}y}{\partial
x^{2}}(l,t)=\frac{\partial^{3}y}{\partial x^{3}}(l,t)=0.
\end{eqnarray}
 Carrying out the conventional Galerkin averaging, we obtain a
set of $n^{th}$ second order ordinary differential equations
coupled. Assuming $n=1$, we obtain the following ordinary
differential equation as the non-dimensionless equation of motion
for the first mode of the beam
\begin{eqnarray}
\label{eq4}
&&\ddot{\chi}(\tau)+2\xi\dot{\chi}+(1+\frac{\Gamma(\tau)}{\Gamma_{cr}})\chi(\tau)+\beta
\chi^3=\gamma_1\chi(\tau-\tau_z)\nonumber\\&&+\gamma_2\dot{\chi}(\tau-\tau_{\dot{z}}),
\end{eqnarray}
 with $\xi=\frac{\lambda}{2\sqrt{I_1\rho SEI}}$, $\beta=-\frac{kI_2}{ESI_1}$, $\Gamma_{cr}=\frac{I_1EI}{I_3}$, $f_0=\frac{4k_ccos(\alpha_c)}{I_1EI}I_4$,
 $\tau_z=\omega_0t_x$,  $\gamma_1=f_0s_1$, $\gamma_2=f_0\omega_0s_2$ and $\omega_0=\sqrt{\frac{EI}{\rho S}I_1}$.\\
 We remind  readers of the following\\
$I_1=\frac{\int_0^{l}\Phi^{IV}(x)\Phi(x)dx}{\int_0^{l}\Phi^{2}(x)dx}$,$I_2=\frac{\int_0^{l}[(\int_0^{l}\Phi^{II}(x)dx)\Phi(x)\Phi^{II}(x)]dx}{\int_0^{l}\Phi^{2}(x)dx}$
$I_3=\frac{\int_0^{l}\Phi^{II}(x)\Phi(x)dx}{\int_0^{l}\Phi^{2}(x)dx}$,
$I_4=\frac{\Phi^{2}(x)}{\int_0^{l}\Phi^{2}(x)dx}$\\ where
$\Phi(x)=-\frac{cosh(k_1l)+cos(k_1l)}{sinh(k_1l)+sin(k_1l)}[(sinh(k_1x)-sin(k_1x))+(cosh(k_1x)-cos(k_1x))]$.\\
To compute these quantities, we utilized Maple and after some algebraic manipulations we have obtained\\
$I_1=\frac{12.36225606}{l^{4}}$, $I_2=\frac{2.362759229}{l^{3}}$,
$I_3=\frac{0.8582473444}{l^{2}}$ and
$I_4=\frac{4.000002152}{l}$.\\
 ${\quad}$${\quad}$ The earthquake signal can be modelled as filtered white noise process while the filter design is
based on a prescribed spectrum of ground motion \cite{R16}. Here, the
Kanai–Tajimi spectral description of the ground motion is used
\begin{eqnarray}
\label{eq5}
S(\omega)=S_0\frac{\omega_g^{4}+4\omega_g^{2}\zeta_g^{2}\omega^{2}}{(\omega^{2}-\omega_g^{2})^{2}+4\omega_g^{2}\zeta_g^{2}\omega^{2}},
\end{eqnarray}
where $\omega_g$; $\zeta_g$ and $S_0$ are parameters which depend
on the  soil characteristics and seismic intensity. The transient
or non-stationary feature of the earthquake is introduced through
an amplitude modulating function \cite{R17}. An equivalent expression
for the evolutionary of earthquake excitation for elastic –
plastic single-degree-of -freedom structures has been presented by \cite{R18,R19,R20}. Where at the first step, the ground
acceleration  is represented as a product of a Fourier series and
an enveloping function as follows
\begin{eqnarray}
\label{eq6} \Gamma(\tau)=e(\tau)[\Sigma_{i=1}^N A_i
cos(\omega_i\tau)+ B_isin(\omega_i \tau)],
\end{eqnarray}
where $A_i$ ;$B_i$ , are $2N$ unknown constants and $\omega_i$,
$i=1 ;2 ;...;N$, are the frequencies embeded in the ground
acceleration $\Gamma$ which are selected such that they span
satisfactory the frequency range $(\omega_0,\omega_c)$. The
function $e(\tau)$ represents the enveloping function that imparts
transient nature to the earthquake acceleration. In the present
study, the envelope function $e(\tau)$ is taken to be given by
\begin{eqnarray}
\label{eq7} e(t)=A_0[exp(-\alpha_1\tau)-exp(-\alpha_2\tau)],
\end{eqnarray}
where $A_0$ $\alpha_1$  and $\alpha_2$  are the parameters for the
enveloping function. The maximum value of the enveloping as per
the above expression is unity. In the present study, for numerical
purposes the frequencies presented in the ground acceleration are
selected such that they span satisfactory the frequency range
$(\omega_0=0.2Hz,\omega_c=25Hz)$. During this analysis $A_0=2.17$,
$\alpha_1=0.13$ and $\alpha_2=0.50$. These choice represents the
earthquake duration to be about 30s, that is typical of the
magnitude 7.0 \cite{R19}. As an illustration the time history of
the optimal ground acceleration and associated Fourier amplitude
spectrum for the earthquake load for this case is shown in Fig. 2
This critical  acceleration will be used for numerical simulation.
 \begin{figure}[!h]
   \centerline{\fbox{\vbox {\hsize0.1cm\hrule width0.1cm height0pt}\includegraphics[width=3.0in]{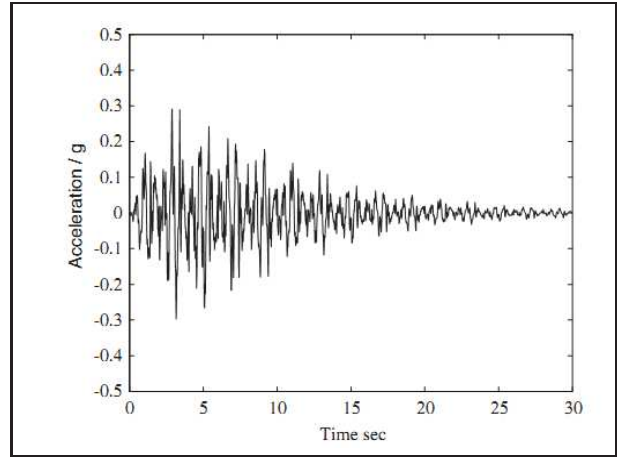}}}
  \caption{Critical acceleration applied to the system}\label{fig.2}
 \end{figure}

 \begin{figure}[!h]
    \centerline{\fbox{\vbox {\hsize0.1cm\hrule width0.1cm height0pt}\includegraphics[width=3.0in]{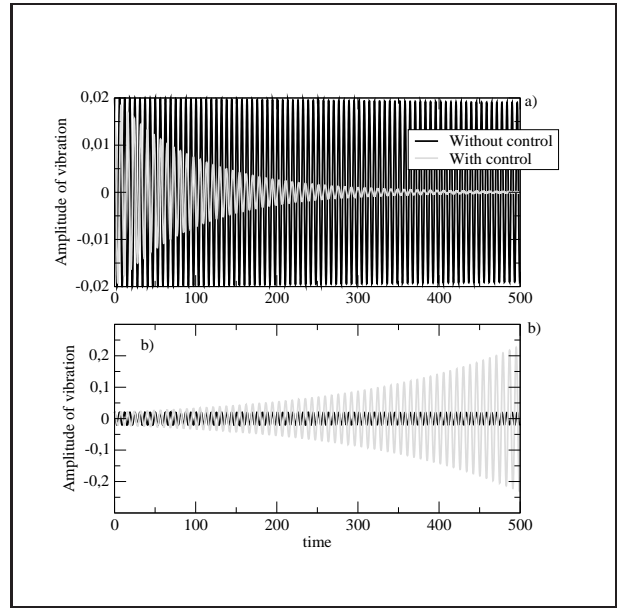}}}
   \caption{Effect of tendon parameters on the stability of the
   control system}\label{fig.3}
  \end{figure}
 \section{Effect of tendon system control on the stability of control design and on the appearance of horseshoes chaos}
 \subsection{On the stability of the control design}
 ${\quad}$${\quad}$ We assume that the tendon system is locked to
 provide the maximum suppression of  oscillations. Then, the
 amplitude dynamics  describes the rate of amplitude change
 depending of the  physical  parameters of the controlled system.
 Taking into all the component acting on the system, the
 non-dimensional governing equation describing the physical system
 under tendon control is given by

 \begin{eqnarray}
 \label{eq8}
 &&\ddot{\chi}(\tau)+2\xi\dot{\chi}+(1+\frac{A_0}{\Gamma_{cr}}[exp(-\alpha_1\tau)-exp(-\alpha_2\tau)]\times\nonumber\\&&[\Sigma_{i=1}^N
 A_i cos(\omega_i\tau)+ B_isin(\omega_i \tau)])\chi(\tau)+\beta
 \chi^3\nonumber\\&&=\gamma_1\chi(\tau-\tau_z)+\gamma_2\dot{\chi}(\tau-\tau_{\dot{z}}).
 \end{eqnarray}
 In the autonomous case the system's stability is explored using
 the Lyapunov concept that examining the fundamental solution
 $e^{St}$ ( $S$ is the Lyapunov exponent). The characteristic
 equation of the eigensystem is then given by
 \begin{eqnarray}
 \label{eq9}
 S^{2}+(2\xi-\gamma_2e^{-St_y})S+1-\gamma_1e^{-St_z}=0,
 \end{eqnarray}
 which is known as a quasipolynomial.
  \begin{figure}[!h]
     \centerline{\fbox{\vbox {\hsize0.1cm\hrule width0.1cm height0pt}\includegraphics[width=3.0in]{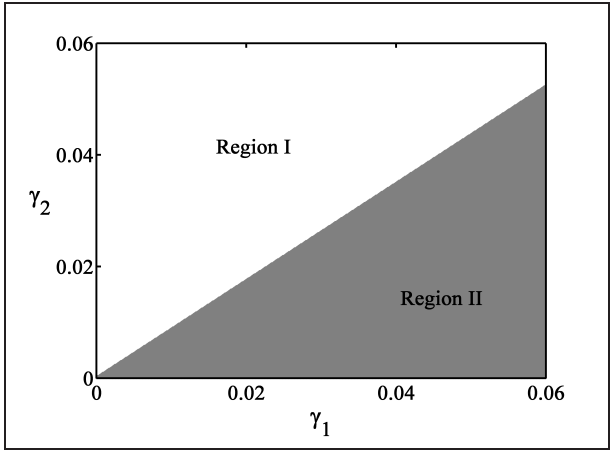}}}
    \caption{Region in space parameters $(\gamma_1,\gamma_2)$
    leading to the suppression of horseshoes chaos (shade region)}\label{fig.4}
   \end{figure}
To study the stability of the controlled system with respect to
the control gain parameter $\gamma_1$ and $\gamma_2$, the method
of D-subdivision is used. The stability boundaries are determined
by the points that yield either  zero root or a pair of pure
imaginary roots of the quasipolynomial \cite{R10}. After analysis
we reach to the conclusion that the control system will remain
stable if the control gain parameters verify these conditions
 $\gamma_1\prec 2\xi$ and
$\gamma_2\prec 1$. Thus, we  come to the fact that the viscolestic
parameters along with time delays have an important effect on the
efficiency of the control process. For illustration we have solved
numerically this equation  using fourth oder Runge kutta
algorithm. We have plotted in figure 3 the evolution of the
amplitude of vibration as a function of time. It is viewed in
figure 3a ( $\gamma_1=0.4$ and $\gamma_2=-0.02$) that the
amplitude decreases as function of time leading to stability while
in figure 3b ( $\gamma_1=0.4$ and $\gamma_2=0.01$) the amplitude
increases with time leading to instability of the control system.
It also appears that as the damping coefficient of the controller
increases the gap between the amplitude of the controlled and
uncontrolled systems decreases, meaning that  as tendon is damped,
the quality of control is destroyed. These findings should be
considered for a better implementation of this concept.  The
analysis of the effect of time delay on the control process give
rise to the conclusion that the stable domain in control space
parameters leading to the efficiency of the control is drastically
reduced as the feedback delay increases \cite{R7,R10}.
 \begin{figure}[!h]
  \fbox{\vbox {\hsize0.1cm\hrule width0.1cm height0.0pt}
      \includegraphics[width=3.0in]{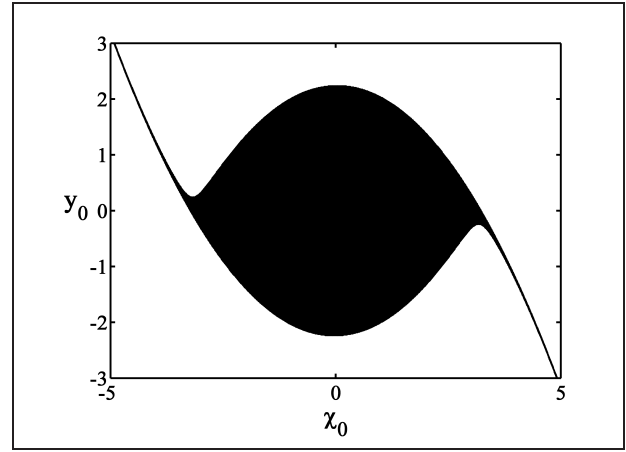}} \\
      \\
      \fbox{\vbox {\hsize0.1cm\hrule width0.1cm height0.0pt}
      \includegraphics[width=3.0in]{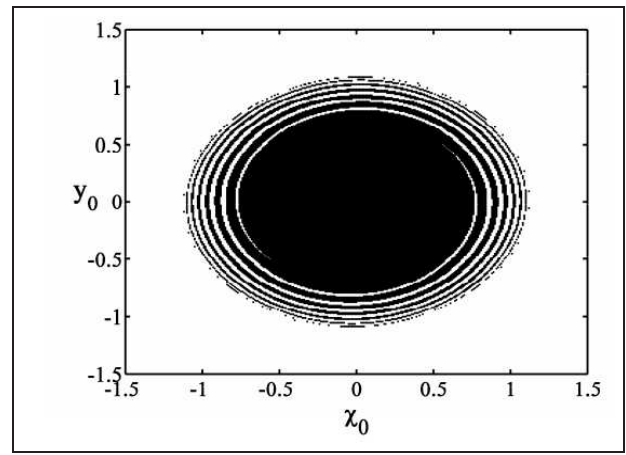}}
     \caption{Basins of attraction for the case a) $\gamma_1=0.05$ and $\gamma_2=0.01$  b)  $\gamma_1=0.02$ and
     $\gamma_2=0.03$}\label{fig.5}
    \end{figure}

 \subsection{On the appearance of horseshoes chaos}
 ${\quad}$${\quad}$ Holmes and Marsden (\cite{R12}) have studied the buckled
 beam subjected to linear damping and periodic transverse forcing .
 They presented a Melnikov-type technique for a class of
 infnite-dimensional systems and gave a criterion under which the
 smale horseshoes chaos appears. In fact, the Melnikov theory has
 been developed to predict the splitting of homoclinic or
 heteroclinic orbits under non-autonomous perturbations. In
 particular, it can be used to establish the existence, or
 non-existence, of transverse homoclinic or heteroclinic  orbits in
 dynamical systems upon adding small non-autonomous terms to the
 governing vector field. Transverse homoclinic or heteroclinic
 orbits, in turn, imply the existence of horseshoes, and therefore
 of chaotic dynamics. The basin of attraction is generally used as
 an indicator for the existence of horseshoes chaos. In this work
 one would like to know how the control strategy affects the
 Melnikov criterion or in what range of the control parameters the
 heteroclinic chaos in our model could be inhibited? To deal with
 such a question, let us express the dynamical structure as
 follows.
 \begin{figure}[!h]
      \centerline{\fbox{\vbox {\hsize0.1cm\hrule width0.1cm height0pt}\includegraphics[width=3.0in]{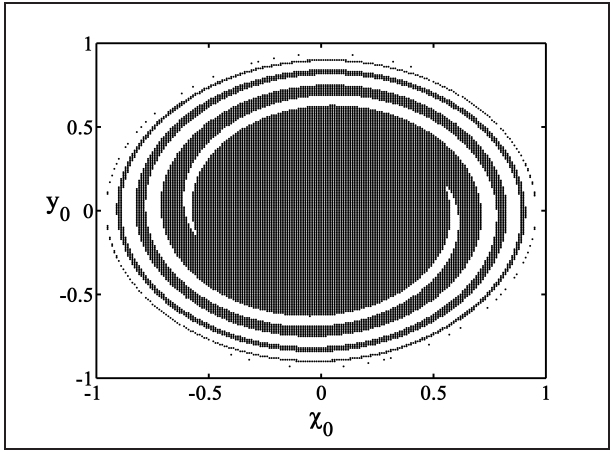}}}
     \caption{Appearance of fractal basin boundary because of
     delay: $\gamma_1=0.05$, $\gamma_2=0.01$ and $\tau=0.9$ }\label{fig.6}
    \end{figure}
 \begin{eqnarray}
 \label{eq10} \dot{U}=F(U)+\varepsilon G(U),
 \end{eqnarray}
 where $U =(\chi; y=\dot{\chi})$ is the state vector, $F=(y;
 -\chi-\beta\chi^{3})$ and $G=(0;
 2\xi\dot{\chi}+(1+\frac{\Gamma(\tau)}{\Gamma_{cr}})\chi(\tau)+\gamma_1\chi(\tau-\tau_z)+\gamma_2\dot{\chi}(\tau-\tau_{\dot{z}}))$.
 The unperturbed system has three fixed points. An hyperbolic fixed
 point  and two elliptic fixed points. It also possesses a
 separatrix solution
 \begin{eqnarray}
 \label{eq11}
 &&\chi_{het}(\tau)=\pm\sqrt{\frac{-1}{\beta}}\tanh(\sqrt{\frac{\tau}{\sqrt{2}}}),\nonumber\\
 &&{{\dot \chi }_{het}}(\tau)=\mp\sqrt{\frac{-1}{2\beta}}sech^{2}(\frac{\tau}{\sqrt{2}}),
 \end{eqnarray}
 known as the separatrix orbit, passing through the hyperbolic
 fixed points. Thus, we can compute the Melnikov function \cite{R11,R12}
 and obtained the following result.
 \begin{eqnarray}
 \label{eq12} &&M(\tau_0)=-2\xi
 J_0+\gamma_1J_{\tau_z}+\gamma_2J_{\tau_y}- \nonumber\\&&\frac{A_0}{\Gamma_{cr}\sqrt{2\beta^{2}}}[\Sigma_{i=1}^{N}(J_{1i}^{A}+J_{1i}^{B})\cos\omega_i\tau_0\nonumber\\&&+\Sigma_{i=1}^{N}(J_{2i}^{A}-J_{2i}^{B})\sin\omega_i\tau_0]
 \end{eqnarray}
 where
 \begin{eqnarray}
 \label{eq13} &&J_0=-\frac{2\sqrt{2}}{3\beta},\nonumber\\
 &&J_{1i}^{A}=2\pi\sqrt{2}A_i[D_i(\alpha_1)+D_i(\alpha_2)], \nonumber\\
 &&J_{2i}^{A}=2\pi\sqrt{2}A_i[E_i(\alpha_1)+E_i(\alpha_2)], \nonumber\\
 &&J_{1i}^{B}=2\pi\sqrt{2}B_i[E_i(\alpha_1)+E_i(\alpha_2)], \nonumber\\
 &&J_{2i}^{B}=2\pi\sqrt{2}B_i[D_i(\alpha_1)+D_i(\alpha_2)], \nonumber\\
 &&J_{\tau_{z}}=\frac{1}{\sqrt{2\beta^{2}}
 \sinh^{2}(\frac{\tau_z}{\sqrt{2}})}[2\tau_z-\sqrt{2}\sinh(\sqrt{2}\tau_z)], \nonumber\\
 &&J_{\tau_{y}}=-\frac{2}{\beta
 \sinh^{3}(\frac{\tau_y}{\sqrt{2}})}[\tau_y\cosh(\frac{\tau_y}{\sqrt{2}})\nonumber\\&&\,\,\,\,\,\,\,\,\,\,\,\,\,\,\,\,-\sqrt{2}\sinh(\frac{\tau_y}{\sqrt{2}})].
 \end{eqnarray}
 with\\
 $D_i(\alpha)=\frac{2\alpha\omega_i\cos[\frac{\pi\alpha}{\sqrt{2}}]\sinh[\frac{\pi\omega_i}{\sqrt{2}}]+(\alpha^{2}-\omega_i^{2})\cosh[\frac{\pi\omega_i}{\sqrt{2}}]\sin[\frac{\pi\alpha}{\sqrt{2}}]}
 {\cos[\pi\alpha\sqrt{2}]-\cosh[\pi\omega_i\sqrt{2}]}$,
 $E_i(\alpha)=\frac{-2\alpha\omega_i\sin[\frac{\pi\alpha}{\sqrt{2}}]\cosh[\frac{\pi\omega_i}{\sqrt{2}}]+(\alpha^{2}-\omega_i^{2})
 \sinh[\frac{\pi\omega_i}{\sqrt{2}}]\cos[\frac{\pi\alpha}{\sqrt{2}}]}{\cos[\pi\alpha\sqrt{2}]-\cosh[\pi\omega_i\sqrt{2}]}$.
 From Eqs. (11), we get the condition for the appearance of the
 Melnikov chaos in the space parameters $(\gamma_1, \gamma_2)$
 which is given by the equation
 \begin{eqnarray}
 \label{eq14}
 &&A_0\sum_{i=1}^{N}(J_{1i}^{A}+J_{1i}^{B}+J_{2i}^{B}-J_{2i}^{A})-\Gamma_{cr}\sqrt{2\beta^{2}}(-2\xi
 J_0\nonumber\\&&+\gamma_1J_{\tau_{z}}+\gamma_2 J_{\tau_{y}})\succ
 0,
 \end{eqnarray}

 Figure 4 displays the domain in parameter space $(\gamma_1,
 \gamma_2)$ leading to the occurrence or suppression of horseshoes
 chaos. We remind readers that this figure is obtained using the
 following dimensionless parameters $\xi=0.0001$ and $\beta=-0.095$
 (here we have considered a steel beam with $S=0.5m^{2}$,
 $E=200MPa$, $\rho=7850kgm^{-3}$, $\lambda=192Nsm^{-1}$ and $L=2m$
 ). The shaded region represents the parameters space for which
 horseshoes chaos cannot appears. It is interesting to verify
 whether the good ranges of control parameter predicted by analytic
 explanation  is  really safe for chaos. To deal with this problem,
 we have looked for the fractality of the basin of attraction by
 solving numerically the base equation.

  Figure 5 displays the basin of attraction for the control gain
  parameters taken in region I and II. These figures confirm the
  previous investigation, since the boundary is regular for the
  control parameter taken in the good region and fractal for the
  other region. Taking into account the delay figure 6 shows
  that the basin where the boundaries where regular, becomes fractal
  because of delays, meaning that the delays can disrupt the control
  strategy.

  \section{Conclusion}
  This paper describes the tendon control strategy in the case of a
  cantilever beam excited by earthquake. The  mathematical modelling
  of the system under control takes into account the fact that,
  through the tendon, the kinetic and potential energy of the system
  is kept small, resulting in small displacements and forces. The
  concept is especially attractive for historical structures because
  their installation requires minimal intervention and can be easily
  removed without causing external visual impact. The earthquake
  model is expanded as a Fourier series, of unknown coefficients,
  that is modulated by an enveloping function. It appears after
  dynamics analysis that forces  in  the tendons that develop during
  the earthquake have a stabilizing effect on the structure.
  Focusing on the occurrence of chaotic dynamics so called
  Horseshoes chaos, the analysis using Melnikov theory  shows the
  effectiveness  of the control strategy presented here in the sense
  that by taking into consideration a selective viscoelastic
  parameter of the tendon system, one can complectly suppress  the
  appearance of horseshoes chaos in the system. Those predictions are
  confirmed and complemented by the numerical simulations from which
  we illustrate the fractal nature of the basins of attraction. It
  is view that the threshold amplitude of earthquake excitation for
  the onset of chaos will move upwards as the physical parameters
  are
  taken in a good region.\\

  {\bf Acknowledgements}\\ B.R. Nana Nbendjo  is grateful to Alexander von Humboldt Foundation for financial
  support within the Georg Forster Fellowship.

\end{document}